\documentclass[twocolumn]{revtex4}

\input{epsf.tex}

\newcommand{\E}{{\cal{E}}}

\renewcommand{\a}{\alpha}

\newcommand{\be}{\begin{equation}}
\newcommand{\ee}{\end{equation}}
\newcommand{\bea}{\begin{eqnarray}}
\newcommand{\eea}{\end{eqnarray}}
\newcommand{\ba}{\begin{array}}
\newcommand{\ea}{\end{array}}
\def\J#1#2#3#4{{#1} {\bf #2}, #3 (#4)}
\def\PRD{Phys. Rev. D}
\def\PR{Phys. Rev.}
\def\PRL{Phys. Rev. Lett.}
\def\RMP{Rev. Mod. Phys.}

\def\LRR{Living Rev. Relativ.}

\def\AJ{Astrophys. J.}
\def\AL{Astron. Lett.}
\def\MNRAS{Mon. Not. R. Astron. Soc.}
\def\JMP{J. Math. Phys.}

\def\CQG{Class. Quantum Grav.}

\def\GRG{Gen. Relativ. Grav.}

\def\ib{{\it ibid.}}

\begin{document}
\draft
\title{Hierarchy of Universal Relations for Neutron Stars\\
in Terms of Multipole Moments}

\author{V.~S.~Manko\,$^\dag$ and E.~Ruiz$\,^\ddag$}
\address{$^\dag$Departamento de F\'\i sica, Centro de Investigaci\'on y de
Estudios Avanzados del IPN, A.P. 14-740, 07000 Ciudad de M\'exico,
Mexico\\$^\ddag$Instituto Universitario de F\'{i}sica Fundamental
y Matem\'aticas, Universidad de Salamanca, 37008 Salamanca, Spain}

\begin{abstract}
Recent studies of the analytical and numerical models of neutron
stars suggest that their exterior field can be described by only
four arbitrary parameters of the 2-soliton solution of Einstein's
equations. Assuming that this is the case, we show that there
exists an infinite hierarchy of the universal relations for
neutron stars in terms of multipole moments that arises as a
series of the degeneration conditions for generic soliton
solutions. Our analysis of the simplest of these relations shows
that the no-hair conjecture for neutron stars proposed by Yagi
{\it et al.} fails to be verified by the perfect fluid models, but
we argue that the conjecture could still be true for the models
involving anisotropic fluid.
\end{abstract}


\maketitle


{\it Introduction.}---In recent years much attention was paid to
the study of the universal properties of neutron stars (NSs) with
the aid of both the numerical and analytical approaches. A
remarkable $I$-Love-$Q$ relation between the NS's moment of
inertia, the tidal Love number and the quadrupole moment was first
dicovered by Yagi and Yunes \cite{YYu} via a numerical analysis of
the Hartle-Thorne slow-rotation approximation \cite{HTh} and then
extended to arbitrary rotation and some new universal properties
by Pappas and Apostolatos \cite{PAp1} and by Chakrabarti {\it et
al.} \cite{CDe}. The exact solutions approach to the analysis of
various phenomena around NSs was introduced by Sibgatullin and
Sunyaev \cite{SSu} who demonstrated that a 3-parameter quadrupole
solution \cite{MMRS} fitted very well the extensive numerical data
of the well-known Cook {\it et al.} paper \cite{CST}; they also
observed that in terms of the dimensionless multipole moments the
properties of NSs independent of the equations of state (EoSs) can
be better seen. Comparison of the analytical and numerical models
of NSs was performed by Berti and Sterligioulas \cite{BSt} with
the aid of the RNS code \cite{SFr,Ste}, and this subsequently led,
via Ryan's method \cite{Rya}, to the revision of multipole moments
in numerical solutions \cite{PAp2}. A better understanding of the
multipole structure of NSs made it possible, on the one hand, to
put the universal relations for NSs into the language of multipole
moments and, on the other hand, to establish \cite{PAp1} that the
above structure is generically determined by only four multipole
moments, thus being universal for all the physically realistic
EoSs known in the literature. Furthermore, in the paper \cite{YKP}
Yagi {\it et al.} conjectured that, similar to black holes, NSs
are likely to verify their own ``no-hair'' theorem according to
which the higher multipoles could be inferred from the form of the
previous four multipoles, and they discussed the numerical
evaluation of the NS's mass-hexadecapole moment in the light of
their conjecture. In \cite{YKP} it was observed in particular that
the 4-parameter 2-soliton solution of Einstein's equations
\cite{MMR,Pap,MRu}, regarded by Pappas and Apostolatos as a
possible analytical model describing the geometry around a
universal NS, possesses a hexadecapole moment whose spin
dependence starts at quadratic order, whereas, according to Yagi
{\it et al.}, this moment should be strictly quartic in angular
momentum.

The objective of the present letter is to demonstrate that the
Yagi {\it et al.} no-hair hypothesis, combined with the
aforementioned 2-soliton solution (henceforth referred to as the
MMR solution), gives rise to a hierarchy of the universal
relations for NSs in terms of multipole moments. The simplest
relation from this hierarchy yields the expression for the
hexadecapole moment $M_4$ of the MMR solution which we will
compare, for two known EoSs, with the empirical formulas of
Ref.~\cite{YKP}.

{\it Multipole moments and the universal relations.}---The
multipole structure of stationary axially symmetric vacuum
spacetimes is well known thanks to the fundamental papers of
Geroch \cite{Ger}, Hansen \cite{Han} and Thorne \cite{Tho}. The
technical calculation of the moments, describing the distributions
of mass and angular momentum, is facilitated by the
Fodor-Hoenselaers-Perj\'es (FHP) procedure \cite{FHP} which makes
use of the Ernst complex potential formalism \cite{Ern} in order
to find the coefficients $m_n$ arising in the expansion of the
function
\be X(z)\equiv z\frac{1-e(z)}{1+e(z)}=
\sum\limits_{n=0}^{\infty}m_nz^{-n} \label{Xz} \ee
when $z\to\infty$. The above $e(z)$ denotes the axis ($\rho=0$)
value of the Ernst complex potential $\E(\rho,z)$ of a particular
solution, $\rho$ and $z$ being the Weyl-Papapetrou coordinates.
The first four quantities $m_n$ coincide with the Geroch-Hansen
(GH) complexified multipole moments $P_n$, $n=\overline{0,3}$,
while other $m_n$, $n\ge4$, are equal to $P_n$ only up to certain
combinations of the lower-order $m_l$, $l<n$ (see Ref.~\cite{FHP}
for the explicit form of those combinations). It is important to
note that the equilibrium models of NSs are described by the
solutions which, in addition to being stationary and axisymmetric,
are also symmetric about the equatorial plane \cite{Ste}, and the
latter symmetry imposes restrictions on the form of the
corresponding axis data $e(z)$ and coefficients $m_n$ in
(\ref{Xz}). As was shown in \cite{Kor,MNe}, the function $e(z)$ of
an equatorially symmetric spacetime satisfies the condition
$e(z)e^{*}(-z)=1$ (the star symbol denotes complex conjugation);
consequently, all even quantities $m_n$ of such a spacetime are
real, and all odd $m_n$ are pure imaginary \cite{Kor}. The same is
true for the corresponding multipoles $P_n$: in the
reflection-symmetric case we have $P_{2k}=M_{2k}$ and
$P_{2k+1}=iJ_{2k+1}$, $k=0,1,\ldots$, where $M_{2k}$ and
$J_{2k+1}$ are, respectively, the mass and angular momentum GH
multipole moments.

Noteworthily, the general class of the extended vacuum soliton
solutions admits parametrization exclusively in terms of the
``multipoles'' $m_n$ \cite{MRu2}:
\bea &&e(z)=\frac{e_-}{e_+}, \quad e_\pm=(L_N)^{-1} \label{ezN} \\
&&\times\left|\begin{array}{cccc} z^N\pm\sum_{n=0}^{N-1}m_n
z^{N-1-n}
& m_N & \ldots & m_{2N-1} \\
z^{N-1}\pm\sum_{n=0}^{N-2}m_n z^{N-2-n}
& m_{N-1} & \ldots & m_{2N-2}  \\
\vdots & \vdots & \ddots & \vdots \\ z\pm m_0 & m_1 & \ldots
& m_N\\
1 & m_0 & \ldots & m_{N-1}
\end{array}\right|, \nonumber \eea
where the $n\times n$ determinant $L_n$ has the form
\be L_n=\left|\begin{array}{cccc} m_{n-1}
& m_n & \ldots & m_{2n-2} \\
m_{n-2}
& m_{n-1} & \ldots & m_{2n-3}  \\
\vdots & \vdots & \ddots & \vdots \\ m_1 & m_2 & \ldots
& m_n\\
m_0 & m_1 & \ldots & m_{n-1}
\end{array}\right|. \label{Ln} \ee
Restricting ourselves to the equatorially symmetric
configurations, we see that the Kerr solution \cite{Ker} is
contained in (\ref{ezN}) as the $N=1$ case, with $m_0=M$ and
$m_1=iJ$, $M$ being the total mass and $J$ the total angular
momentum \cite{Kom}. The next, $N=2$ specialization of formulas
(\ref{ezN}), determines the MMR solution that was recently
regarded and advocated as describing the exterior field of a
universal NS \cite{PAp1,Pap,PAp3}; it has four arbitrary
parameters corresponding to four arbitrary multipole moments:
\bea &&m_0=M_0\equiv M, \quad m_1=iJ_1\equiv iJ, \nonumber\\
&&m_2=M_2, \quad m_3=iJ_3, \label{MMR} \eea
where $M_2$ is the mass quadrupole moment and $J_3$ is the angular
momentum octupole moment (the explicit form of the MMR solution in
two different parametrizations can be found in Ref.~\cite{MRu}).
If we now assume that the results of Pappas and Apostolatos
\cite{PAp1} obtained on the basis of a variety of very convincing
arguments are correct and the geometry around NSs is indeed
determined by only four multipole moments (\ref{MMR}), then,
bearing in mind the no-hair hypothesis for NSs put forward by Yagi
{\it et al.} \cite{YKP}, we inevitably arrive at the MMR spacetime
as the simplest and hence most suitable model for the exterior of
a NS complying with the conditions of papers \cite{PAp1,YKP}. The
fact that the MMR solution is the simplest one possessing the
required four moments is very important in itself because it makes
this solution in a sense similar to the Kerr spacetime whose
unique property is that it is the simplest possible solution among
infinite number of the 2-parameter solutions defined by the
parameters of mass and angular momentum. Clearly, the higher GH
multiple moments of the MMK solution will then be some
well-defined functions of the above four parameters that can be
found from the corresponding axis data by means of the FHP
procedure.

The explicit expressions of the multipoles $M_{2n}$ and
$J_{2n+1}$, $n\ge2$, as functions of the moments (\ref{MMR}), in
the case of the MMR solution would give us the simplest hierarchy
of the universal relations for NSs. Obviously, each relation from
this hierarchy determines how the higher multipole $M_{2n}$ or
$J_{2n+1}$, with a specific $n$, depends on the first four lower
moments (\ref{MMR}); however, since such relations involve only
one higher multipole, they do not actually provide any information
about possible interrelations between the higher multipole moments
themselves. So it is remarkable that there does exist a more
sophisticated hierarchy of the universal relations for NSs that
directly connect different higher multipoles with each other. This
new hierarchy arises in (\ref{ezN}) as a series of the
degeneration conditions of the solitonic solutions with $N>2$ to
the $N=2$ case. Indeed, as was shown in \cite{MRu2}, the general
$N$-soliton solution degenerates to the $(N-1)$-soliton case when
the determinant $L_N$ defined by (\ref{Ln}) becomes equal to zero;
then further degeneration would require zero values of the
determinants $L_{N-1}$, $L_{N-2}$ and so on, until we finally
arrive at the 2-soliton solution by means of the conditions
$L_3=0$, $L_2\ne0$, the latter nonequality being needed to stop
the degeneration process. By inverting this reasoning, we can say
that the higher multipole moments of the MMR 2-soliton solution
must be such that the conditions
\be L_n=0 \quad \mbox{for all} \quad n>2 \label{cond2s} \ee
are satisfied. It is easy to see from (\ref{Ln}) that the above
conditions (\ref{cond2s}) establish how the moments $m_{2n-2}$
depend on the moments $m_{2n-3}$, or, roughly speaking and taking
into account the equatorial symmetry of the MMR solution, how the
GH mass multipoles $M_{2n-2}$ depend on the spin multipoles
$J_{2n-3}$.

The simplest of the relations (\ref{cond2s}), accounting for
(\ref{MMR}), takes the form
\be L_3=\left|
\begin{array}{ccc}
M_2 & iJ_3 & m_4 \\ iJ & M_2 & iJ_3 \\
 M & iJ & M_2
\end{array}
\right|=0, \label{L3c} \ee
whence we get
\be m_4=\frac{M_2^3+2JJ_3M_2-MJ_3^2}{MM_2+J^2}. \label{m4} \ee

This formula is of importance because it permits us to compare the
dependence of the hexadecapole moment $M_4$ on the angular
momentum $J$ in the MMR solution and in the known numerical models
for NSs. In the paper \cite{YKP} it was found, with the aid of the
quartic-order slow-rotation approximation and numerical solutions,
that similar to the Kerr spacetime, the multipoles $M_2$, $J_3$
and $M_4$ of NSs are proportional, respectively, to $J^2$, $J^3$
and $J^4$, so that, according to Yagi {\it et al.}, the
hexadecapole moment $M_4$ should not contain any term proportional
to $J^2$. Supposing that $M_2\propto J^2$ and $J_3\propto J^3$,
one can readily see that the quantity $m_4$ in (\ref{m4}) is
proportional to $J^4$. Nonetheless, the relation of $m_4$ to the
GH hexadecapole moment $M_4$ is defined by the formula \cite{FHP}
\be m_4=M_4+\frac{1}{7}M(J^2+MM_2), \label{mM4} \ee
and it is clear that the second term on the right-hand side of
(\ref{mM4}) is proportional to $J^2$, so that the expression of
$M_4$ of the MMR solution necessarily contains terms proportional
both to $J^2$ and $J^4$. Moreover, the condition for $M_4$ to be
proportional strictly to $J^4$ implies $M_2=-J^2/M$, which is
exactly the value of the mass-quadrupole moment of the Kerr
solution. Mention that the situation will be the same if we opt to
use the multipole moments constructed according to Thorne's
definition \cite{Tho}, since these are known \cite{Gur} to be
proportional to the GH multipoles. Therefore, the structure of the
mass hexadecapole moment in the MMR solution and in the solutions
analyzed by Yagi {\it et al.} is clearly different, and this
difference is seemingly determined by the specific properties of
the interior solutions in the approximate and numerical models
considered in \cite{YKP}. We shall return to this point later on.

As was remarked in \cite{PAp1}, the universal relations for NSs
must be independent of the total mass $M$ when these are rewritten
in terms of the rescaled, dimensionless moments. Then, bearing
this in mind and introducing the rescaled moments via the formulas
\be J=jM^2, \quad M_2=qM^3, \quad J_3=sM^4, \quad M_4=\mu M^5,
\label{mdim} \ee
it is possible, taking into account (\ref{mM4}), to rewrite
formula (\ref{m4}) in the `$M$ free' form
\be \mu=-\frac{1}{7}(j^2+q)+\frac{q^3+2jqs-s^2}{j^2+q},
\label{m4u} \ee
thus demonstrating that the relation $L_3=0$ safely passes the
additional test of universality.

Remarkably, the next relation from the hierarchy (\ref{cond2s}),
$L_4=0$, which involves the multipoles $M_6$ and $J_5$, on account
of (\ref{m4}) becomes independent of $M_6$ and yields directly the
expression for the spin multipole $J_5$; written in terms of the
dimensionless moments, with $J_5=\chi M^6$, it takes the form
\bea \chi=\frac{1}{21}(j^3+8jq-7s)+\frac{js^2-q^2s}{j^2+q}
\nonumber\\-\frac{jq^4-3jqs^2-3q^3s+s^3}{(j^2+q)^2}. \label{m5u}
\eea

In principle, it is not difficult to show that, after an
appropriate rescaling, $M$ always cancels out from the generic
relation $L_n=0$ in the equatorially symmetric case under
consideration.

{\it An application.}---The fact that the mass hexadecapole moment
of the MMR solution is not strictly quartic in spin does not
actually lead immediately to the conclusion that it contradicts
the Yagi {\it et al.} analysis, because $\mu$ in (\ref{mdim}) and
(\ref{m4u}) can be always formally put into the form
$\mu=\a_0j^4$, permitting a trivial estimation of the coefficient
$\a_0$ for any concrete model of a NS. Therefore, for being able
to draw a more substantiated conclusion, it is yet necessary to
compare the values of the moment $M_4$ calculated, for some
available numerical models of NSs, with the aid of the Yagi {\it
et al.} approach \cite{YKP} based on Ryan's method \cite{Rya}, on
the one hand, and by means of our formula (\ref{m4}) after the
substitution in it of (\ref{mM4}), on the other hand. To obtain
the values of the first type, which we denote as $M_4^{(n)}$, we
used the hints left in \cite{YKP} for the evaluation of $M_4$ in
the case of EoSs AU and L; the concrete models (originally due to
Berti and Stergioulas \cite{BSt}) were taken from tables II and VI
of the Supplement to the paper \cite{PAp1}. For the same models we
then estimated the mass-hexadecapole moments $M_4^{(a)}$ using our
analytical formula. The results are summarized in Tables~I and II,
which also include the ratios $M_4^{(a)}/M_4^{(n)}$ for
convenience. One can see that the correspondence between
$M_4^{(n)}$ and $M_4^{(a)}$ for all three sequences of EoS AU from
Table~I is quite reasonable, though ranging from almost full
coincidence to an appreciable difference. As for Table~II, it
seems that the disagreement between the values $M_4^{(n)}$ and
$M_4^{(a)}$ calculated for the models with the EoS~L is rather
substantial for all instances, whence we tentatively conclude that
the latter EoS is probably not appropriate for modeling the
interior of NSs. In this respect, it would be interesting to
perform a similar comparative analysis of the mass-moment $M_4$
for other known EoSs since, as we are convinced, a good exterior
solution for NSs should be able not only to match routinely any
kind of numerical model independently of its real value, but more
importantly must help to distinguish among good and bad numerical
solutions or EoSs.

An effort to explain the discrepancy between the values
$M_4^{(n)}$ and $M_4^{(a)}$ in Table~I leads to several far-going
conclusions. Let us fist note that although at the beginning we
were attributing that discrepancy to a misinterpretation of the
higher GH multipoles by Yagi {\it et al.}, this is not really the
case, as we have been able to clarify recently with the aid of the
authors of \cite{YKP}. The essence of the problem seems to lie in
the fact that the approximate and numerical interior solutions for
NSs analyzed in \cite{YKP} are developed on the basis of the
perfect fluid models, thus being globally inconsistent with the
possible sources of the MMR metric which must involve the
anisotropic fluid. Indeed, it is well known \cite{WNe} that
perfect fluid cannot be a source for the Kerr spacetime, and a
recent paper of Hernandez-Pastora and Herrera \cite{HPH} gives a
nice example of the physically meaningful anisotropic fluid
solution matching the Kerr metric on the surface of zero pressure.
Therefore, the MMR solution, being a generalization of the Kerr
spacetime, must also have anisotropic fluid as its source, which
means that even though this solution can match perfectly well any
numerical model up to the spin-octupole moment $J_3$ thanks to the
four arbitrary parameters it possesses, the discrepancy will still
show up in the higher multipole moments, being greater (smaller)
the greater (smaller) is the deviation of the interior solution of
a concrete numerical model from the anisotropic fluid source of
the MMR metric. Moreover, since NSs can in principle collapse into
a black hole described by the Kerr solution, it would be plausible
to infer that the interior solution of a generic NS model, similar
to Kerr's source, must also take into account the anisotropy in
the pressure, the importance of which for rotating bodies was
highlighted and substantiated in the paper \cite{HPH}. As a
result, it is highly improbable that the Yagi {\it et al.} no-hair
conjecture for NSs could be verified by the approximate and
numerical solutions involving perfect fluid as the interior of a
Ns. On the other hand, the Yagi {\it et al.} conjecture is fully
supported by the MMR solution whose higher multipoles are all
determined by the lower four moments in a well defined form, and
in the future it only remains to complement this solution with the
anisotropic fluid models of a new generation congruent with the
MMR interior, thus giving rise to the `no-hair' global solutions
of NSs.

It is worth noting in conclusion that the scope of applicability
of the universal relations considered in the present letter
actually goes beyond the NSs only. Thus, for instance, the whole
hierarchy (\ref{cond2s}) is eligible in the case of the Kerr
solution too, and besides should be supplemented with the relation
$L_2=0$ $\Leftrightarrow$ $M_2=-J^2/M$ which shows that a NS
collapses to a black hole when its quadruple moment becomes that
of the Kerr solution, independently of the value of its
spin-octupole moment $J_3$. Moreover, other stellar objects with a
richer structure than that of NSs and hence requiring more than
four arbitrary real parameters for its description, could be
analytically approximated by the $N=3,4,\ldots$ extended soliton
solutions, in which case the inequality in (\ref{cond2s}) starts,
respectively, from $3,4,\ldots$ The degeneration conditions then
would reflect in particular the evolution of stars from one type
to another.

\section*{Acknowledgments}

We are indebted to Kent Yagi, Georgios Pappas, Nicolas Yunes and
Theocharis Apostolatos for discussing with us the details of their
method. We are also thankful to the referees for valuable remarks
that helped us to considerably improve the presentation. This work
was partially supported by the CONACYT of Mexico, and by Project
FIS2015-65140-P (MINECO/FEDER) of Spain.

\begin{table}[htb]
\caption{Multipole moments for three sequences of
Berti-Stergioulas numerical models constructed with EoS AU from
table~II of \cite{PAp2}. The procedure for evaluating $M_4^{(n)}$
is given in Ref.~\cite{YKP}, while the values of $M_4^{(a)}$ have
been obtained with the aid of formula (\ref{m4}) of the present
paper.}
\begin{center}
\begin{tabular}{lllllll}
\hline \hline $M$ \hspace{1cm} & $J/M^2$ \hspace{0.5cm} & $M_2$
\hspace{0.5cm} &
$J_3$ \hspace{1cm} & $M_4^{(n)}$ \hspace{0.5cm} & $M_4^{(a)}$ \hspace{0.7cm} & $M_4^{(a)}/M_4^{(n)}$ \\
\hline 2.072
& 0.201  & -1.45 & -1.14 & 2.217 & 1.947 & 0.878 \\
2.087  & 0.414  & -6.08 & -10.0 & 40.21 & 25.51 & 0.634 \\
2.097  & 0.529  & -9.96 & -21.2 & 107.7 & 71.38 & 0.662 \\
2.108 & 0.616 & -13.6 & -34.1 & 199.1 & 121.0 & 0.608 \\
2.112  & 0.661 & -15.7 & -42.7 & 264.4 & 161.4 & 0.610 \\ \hline
3.164 & 0.194 & -1.68 & -1.37 & 2.202 & 1.825 & 0.829 \\
3.207 & 0.406 & -8.08 & -14.5 & 41.97 & 30.23 & 0.720 \\
3.253 & 0.550 & -16.1 & -41.4 & 140.3 & 116.2 & 0.828 \\
3.291 & 0.645 & -23.9 & -75.1 & 263.7 & 253.1 & 0.960 \\
3.318 & 0.706 & -30.3 & -107. & 376.7 & 403.0 & 1.07 \\ \hline
3.388 & 0.510 & -12.9 & -28.1 & 101.2 & 65.81 & 0.650 \\
3.393 & 0.520 & -13.7 & -31.1 & 109.2 & 75.78 & 0.694 \\
3.422 & 0.587 & -19.1 & -51.6 & 176.3 & 148.6 & 0.843 \\
3.458 & 0.659 & -26.0 & -82.7 & 277.9 & 277.6 & 0.999 \\
3.487 & 0.713 & -32.5 & -115. & 378.4 & 428.8 & 1.133 \\ \hline
\hline
\end{tabular}
\end{center}
\end{table}

\begin{table}[htb]
\caption{Multipole moments for three sequences of
Berti-Stergioulas numerical models constructed with EoS L from
table~VI of \cite{PAp2}.}
\begin{center}
\begin{tabular}{lllllll}
\hline \hline $M$ \hspace{1cm} & $J/M^2$ \hspace{0.5cm} & $M_2$
\hspace{0.5cm} & $J_3$
\hspace{1cm} & $M_4^{(n)}$ \hspace{0.5cm} & $M_4^{(a)}$ \hspace{0.7cm} & $M_4^{(a)}/M_4^{(n)}$ \\
\hline 2.071
& 0.194  & -2.76 & -2.28 & 7.405 & 5.730 & 0.774 \\
2.080 & 0.417  & -12.2 & -22.0 & 159.3 & 90.39 & 0.567 \\
2.087 & 0.543 & -20.3 & -47.9 & 460.9 & 243.6 & 0.529 \\
2.095 & 0.650 & -28.6 & -81.3 & 952.9 & 478.5 & 0.502 \\
2.097 & 0.698 & -32.9 & -100. & 1269. & 628.6 & 0.495 \\ \hline
4.012 & 0.178 & -3.80 & -4.23 & 16.8 & 8.943 & 0.532 \\
4.051 & 0.375 & -18.5 & -45.4 & 338.5 & 138.6 & 0.410 \\
4.098 & 0.528 & -40.3 & -144. & 1367. & 599.8 & 0.439 \\
4.139 & 0.635 & -62.6 & -279. & 2930. & 1415. & 0.483 \\
4.167 & 0.700 & -79.8 & -401. & 4399. & 2276. & 0.518 \\ \hline
4.321 & 0.479 & -29.5 & -90.2 & 1059. & 308.0 & 0.291 \\
4.325 & 0.489 & -31.9 & -101. & 1153. & 358.4 & 0.311 \\
4.355 & 0.555 & -45.2 & -170. & 1950. & 709.4 & 0.364 \\
4.396 & 0.641 & -66.0 & -299. & 3560. & 1488. & 0.418 \\
4.420 & 0.686 & -79.4 & -394. & 4742 & 2140. & 0.451 \\ \hline
\hline
\end{tabular}
\end{center}
\end{table}


\begin{references}

\bibitem{YYu} K. Yagi and N. Yunes, Science
{\bf 341}, 365 (2013); \J{\PRD}{88}{023009}{2013}.

\bibitem{HTh} J. B. Hartle and K. S. Thorne, \J{\AJ}{153}{807}{1968}.

\bibitem{PAp1} G. Pappas and T. A. Apostolatos, \J{\PRL}{112}{121101}{2014}.

\bibitem{CDe}  S. Chakrabarti, T. Delsate, N. Gurlebeck,
and J. Steinhoff, \J{\PRL}{112}{201102}{2014}.

\bibitem{SSu} N. R. Sibgatullin and R. A. Sunyaev,
\J{\AL}{24}{774}{1998}; \J{\ib}{26}{699}{2000}.

\bibitem{MMRS} V. S. Manko, J. Mart\'in, E. Ruiz, N. R. Sibgatullin,
and M. N. Zaripov, \J{\PRD}{49}{5144}{1994}.

\bibitem{CST} G. B. Cook, S. L. Shapiro, and S. A. Teukolsky,
\J{\AJ}{424}{823}{1994}.

\bibitem{BSt} E. Berti and N. Stergioulas, \J{\MNRAS}{350}{1416}{2004}.

\bibitem{SFr} N. Stergioulas and J. L. Friedman, \J{\AJ}{444}{306}{1995}.

\bibitem{Ste} N. Stergioulas, \J{\LRR}{6}{3}{2003}.

\bibitem{Rya} F. D. Ryan, \J{\PRD}{52}{5707}{1995}, \J{\ib}{55}{6081}{1997}.

\bibitem{PAp2} G. Pappas and T. A. Apostolatos, \J{\PRL}{108}{231104}{2012}.

\bibitem{YKP} K. Yagi, K. Kyutoku, G. Pappas, N. Yunes, and T. A.
Apostolatos, \J{\PRD}{89}{124013}{2014}.

\bibitem{MMR} V. S. Manko, J. Mart\'in, and E. Ruiz,
\J{\JMP}{36}{3063}{1995}.

\bibitem{Pap} G. Pappas, \J{\MNRAS}{422}{2581}{2012}.

\bibitem{MRu} V. S. Manko and E. Ruiz, \J{\PRD}{93}{104051}{2016}.

\bibitem{Ger} R. Geroch, \J{\JMP}{13}{394}{1972}.

\bibitem{Han} R. O. Hansen, \J{\JMP}{15}{46}{1974}.

\bibitem{Tho} K. S. Thorne, \J{\RMP}{52}{299}{1980}.

\bibitem{FHP} G. Fodor, C. Hoenselaers, and Z. Perj\'es, \J{\JMP}{30}{2252}{1989}.

\bibitem{Ern} F. J. Ernst, \J{\PR}{167}{1175}{1968}.

\bibitem{Kor} P. Kordas, \J{\CQG}{12}{2037}{1995}.

\bibitem{MNe} R. Meinel and G. Neugebauer, \J{\CQG}{12}{2045}{1995}.

\bibitem{MRu2} V. S. Manko and E. Ruiz, \J{\CQG}{15}{2007}{1998}.

\bibitem{Ker} R. P. Kerr, \J{\PRL}{11}{237}{1963}.

\bibitem{Kom} A.~Komar, \J{\PR}{79}{084024}{2009}.

\bibitem{PAp3} G. Pappas and T. A. Apostolatos, \J{\MNRAS}{429}{3007}{2013}.

\bibitem{MTW} C. W. Misner, K. S. Thorne, and J. A. Wheeler,
Gravitation (W. H. Freeman, San Francisco, 1973).

\bibitem{Gur} Y. G\"ursel, \J{\GRG}{15}{737}{1983}.

\bibitem{WNe} T. Wolf and G. Neugebauer, \J{\CQG}{9}{L37}{1992}.

\bibitem{HPH} J. L. Hernandez-Pastora and L. Herrera, \J{\PRD}{95}{024003}{2017}.






\end{references}
\end{document}